\documentclass[twocolumn,aps,superscriptaddress,nofootinbib,floatfix,linenumbers]{revtex4}
\usepackage{amsfonts}
\usepackage{float}
\usepackage{placeins}
\usepackage{graphicx}
\usepackage[hidelinks]{hyperref}
\usepackage{color,amsmath,amssymb,bm}
\usepackage{tensor}

\usepackage[normalem]{ulem}  

\renewcommand{\sout}{\bgroup \color{black} \ULdepth=-.5ex \ULset}

\def\blfootnote{\xdef\@thefnmark{}\@footnotetext}

\newcommand{\beq}{\begin{equation}}
\newcommand{\eeq}{\end{equation}}
\newcommand{\bea}{\begin{eqnarray}}
\newcommand{\eea}{\end{eqnarray}}

\begin{document}

\title{Event shape Engineering analysis of D meson in ultrarelativistic heavy ion collisions }

\author{Maria Lucia Sambataro}
\affiliation{Department of Physics and Astronomy, University of Catania, Via S. Sofia 64, 1-95125 Catania, Italy}
\affiliation{Laboratori Nazionali del Sud, INFN-LNS, Via S. Sofia 62, I-95123 Catania, Italy}

\author{Yifeng Sun}
\affiliation{School of Physics and Astronomy, Shanghai Key Laboratory for Particle Physics and Cosmology, and Key Laboratory for Particle Astrophysics and Cosmology (MOE), Shanghai Jiao Tong University, Shanghai 200240, China}
\affiliation{Department of Physics and Astronomy, University of Catania, Via S. Sofia 64, 1-95125 Catania, Italy}
\affiliation{Laboratori Nazionali del Sud, INFN-LNS, Via S. Sofia 62, I-95123 Catania, Italy}

\author{Vincenzo Minissale}
\affiliation{Department of Physics and Astronomy, University of Catania, Via S. Sofia 64, 1-95125 Catania, Italy}
\affiliation{Laboratori Nazionali del Sud, INFN-LNS, Via S. Sofia 62, I-95123 Catania, Italy}

\author{Salvatore Plumari}
\email{salvatore.plumari@dfa.unict.it}
\affiliation{Department of Physics and Astronomy, University of Catania, Via S. Sofia 64, 1-95125 Catania, Italy}
\affiliation{Laboratori Nazionali del Sud, INFN-LNS, Via S. Sofia 62, I-95123 Catania, Italy}

\author{Vincenzo Greco}
\affiliation{Department of Physics and Astronomy, University of Catania, Via S. Sofia 64, 1-95125 Catania, Italy}
\affiliation{Laboratori Nazionali del Sud, INFN-LNS, Via S. Sofia 62, I-95123 Catania, Italy}

\date{\today}

\begin{abstract}
We describe the propagation of charm quarks in the quark-gluon plasma (QGP) by means of an event-by-event transport approach.  In our calculations the non-perturbative interaction between heavy quarks and light quarks has been taken into account through a quasi-particle approach with thermal light quark masses tuned to reproduce lQCD thermodynamics.
We found that the flow observables $v_2$ and $v_3$ of D mesons are comparable with the experimental measurements for Pb+Pb collisions at 5.02 ATeV in different ranges of centrality selections. The results are analyzed with Event-Shape Engineering technique. The comparison of the anisotropic flow coefficients $v_n$ with experimental data show a quite well agreement with experimental data for different flow vector $q_2$ selections, which confirms the strong coupling between charm quarks and light quarks in the QCD matter.
Furthermore, we present here a novel study of the event-by-event correlations between flow harmonics of $D$ mesons and soft hadrons at LHC energy with the Event-Shape Engineering technique that can put further constraints on heavy quark transport coefficients toward a solid comparison between the phenomenological determination and the lattice QCD calculations.
\end{abstract}

\maketitle

\section{Introduction}

Heavy quarks (HQs), mainly charm and bottom quarks, are produced in the early stages of a ultra-relativistic Heavy Ion collision (uRHIC) by hard processes described by perturbative QCD calculations. They have been identified as one among the few probes 
which may allow for a direct study of the QGP properties. This is manly due to the following reasons related to their large masses: i) they have a very short formation time ($\tau \sim 0.1$ fm/c), ii) HQs are not expected to reach a full thermalization. 
Therefore, they can probe the entire space-time evolution of the matter created in these collisions. Furthermore, due to their interaction with the medium produced (quarks and gluons) they can keep this information in their final state as constituent of heavy hadrons
mainly $D$, $B$ mesons and $\Lambda_c$, $\Lambda_b$ baryons.  Among the several observables studied in uRHICs two of them have been long
investigated for HF hadrons: the heavy mesons nuclear modification factor  $R_{AA}(p_T)$ \cite{STAR:2006btx,PHENIX:2005nhb,ALICE:2015vxz}, and the so called elliptic flow, $v_2(p_T)$. 
The first gives an information about the change of the spectrum in $AA$ collision with respect to a simple
$pp$ superposition. 
The second gives a measure of the anisotropy in the particle angular distribution and allows to investigate the coupling of the HQs with the medium and their participation in the collective motion. 
Both these quantity have been extensively used as probe of the  Quark Gluon Plasma ~\cite{PHENIX:2006iih,ALICE:2014qvj}.
From a theoretical point of view, in recent years, several studies have done efforts in the direction of studying both of these observables to understand the heavy quark dynamics in QGP ~\cite{vanHees:2005wb,vanHees:2007me,Gossiaux:2008jv,Das:2009vy,Alberico:2011zy,Uphoff:2012gb,Lang:2012nqy,Song:2015sfa,Song:2015ykw,Das:2013kea,Cao:2015hia,Das:2015ana,Cao:2017hhk,Das:2017dsh,Sun:2019fud,Cao:2018ews,Rapp:2018qla}.\\
More recent measurements on the $D$ mesons and $\Lambda_c$ production in nuclear collisions unfolded the 
possibility to get access to richer information. In particular, 
there have been studies that have opened the way 
to start to address the issue of medium-modification of heavy-flavour hadronization
making possible to get information about heavy-quark hadronization and possibly to validate 
models based on the recombination \cite{Plumari:2017ntm},
and recently these studies  
have been extended to the case of small collision systems like proton-nucleus
and proton-proton collision 
providing insight on 
the heavy flavor hadronization 
mechanism and
the possibility that also in such small systems the QGP can be formed \cite{Minissale:2020bif}.\\
More information about the interactions of heavy quarks with the medium can be obtained through measurements 
of the azimuthal distributions of heavy-flavour hadrons produced in a HIC.  
Because of the interaction of heavy quarks with the medium, the spatial anisotropy in the early stages of nucleus–nucleus collisions is converted into an azimuthally anisotropic distribution in momentum space for the final particles produced from the QGP. This anisotropy is characterised in terms of the Fourier coefficients $v_n$ of the final distribution respect to the symmetry-plane angles $\Psi_n$ (for the harmonic of order $n$). The second coefficient of this expansion is the elliptic flow $v_2$ which is the dominant contribution in non-central collisions.
The higher order harmonics usually studied are the triangular flow $v_3$, and the $v_4 $; the measurement of odd harmonics of heavy-flavour hadrons or their event-by-event fluctuations coming from the random nucleon positions can provide a richer information on the initial conditions. Recently, the triangular flow $v_3$ has been investigated in studies based on Langevin or Boltzmann approaches on an event-by-event analysis ~\cite{Nahrgang:2014vza,Prado:2016szr,Beraudo:2018tpr,Katz:2019fkc,Plumari:2019hzp}.
One of the new tools that can be used  to further investigate the dynamics of
heavy quarks in the medium is the Event Shape Engineering (ESE) technique \cite{Schukraft:2012ah}. 
With this procedure it is possible to select events in heavy-ion collisions 
based on the geometry of the fireball, enabling to select sub-samples of events characterized by an 
high or low eccentricity, furnishing the chance to study the system evolution dynamics
and the role of the initial conditions.
This technique was first used by the ALICE collaboration in the light-flavour sector to
study the interplay between the initial geometry of the nucleus–nucleus collisions and the subsequent
evolution of the system \cite{ALICE:2015lib} and recently it was extended to investigate the sensitivity of the D meson $v_2$ to
the event-shape selection \cite{ALICE:2018gif}. These measurements have shown a sizeable sensitiveness to
the event-shape selection, confirming a correlation between the D-meson azimuthal anisotropy and
the underlying collective expansion of the bulk matter.
In this paper we use an event-by-event transport approach for the evolution of heavy quarks and we perform an analysis of the heavy-quark production using the ESE technique, in order to study 
the correlation of heavy-quark anisotropic flows $v_n-v_m$ correlations.

The paper is organized as follows. In section II we discuss briefly the Boltzmann transport approach 
used to describe the HQs evolution and the   hadronization model used, furthermore we will describe the initial 
conditions employed in the simulation. In sections III we discuss the comparison between the simulation 
results and the experimental data for the anisotropic flows coefficients $v_n$ at different 
centralities. In section IV we introduce the ESE technique and we show and discuss the outcome for the 
anisotropic flows coefficients and their correlations. Finally, section V contains a summary and some 
concluding remarks.

\section{Heavy quarks transport approach}
We use a transport code developed to perform studies of the dynamics of relativistic heavy-ion collisions at both RHIC and LHC energies \cite{Plumari:2012ep, Ruggieri:2013bda, Scardina:2012mik, Ruggieri:2013ova, Puglisi:2014sha,  Scardina:2014gxa, Plumari:2015sia, Plumari:2015cfa, Scardina:2017ipo, Plumari:2019gwq, Sun:2019gxg}.
The space-time evolution of gluons and light quarks as well as of charm quarks distribution functions is described by mean of the Relativistic Boltzmann Transport (RBT) equations given by
\begin{eqnarray}
& & p^{\mu}_q \partial_{\mu}f_{q}(x,p)= {\cal C}[f_q,f_g](x_q,p_q) \label{eq:RBTeqq} \\
& & p^{\mu}_g \partial_{\mu}f_{g}(x,p)= {\cal C}[f_q,f_g](x_g,p_g) \label{eq:RBTeqg} \\
& & p^{\mu} \partial_{\mu}f_{Q}(x,p)= {\cal C}[f_q,f_g,f_{Q}](x,p) \label{eq:RBTeqQ}
\label{B_E} 
\end{eqnarray}
where $f_i(x,p)$ is the on-shell phase space one-body distribution function for the $i-$th parton and in the right-hand side ${\cal{C}}[f_q, f_g, f_{Q}](x,p)$ is the relativistic Boltzmann-like collision integral accounting for $2\rightarrow2$ scattering processes.
For the quarks and gluons distribution functions that compound the QGP medium, the evolution is described by Eqs.~\eqref{eq:RBTeqq} and ~\eqref{eq:RBTeqg} in this paper. We assume that these Eqs. are independent of $f_{Q}$, which correspond to discard collisions between charm quarks and the bulk in the determination of light quarks and gluons distribution functions, which is by far a solid approximation. In both collision integrals the total cross section is determined in order to keep the ratio $\eta/s=1/(4\pi)$  fixed during the evolution of the QGP, see Refs~\cite{Plumari:2019gwq, Plumari:2015cfa, Ruggieri:2013ova, Plumari:2012xz} for more details.
Furthermore, in the present paper we have employed a bulk with thermal massive quarks and gluons according to a Quasi-Particle Model (QPM) such that the system has approximately the lattice QCD equation of state, with a softening of the equation of state  consistent with a decreasing speed of sound approaching the cross-over region \cite{Borsanyi:2010cj}.
For the heavy quark interacting with the bulk medium we consider a QPM whose main feature is that the resulting coupling is significantly stronger than the one coming from pQCD running coupling, particularly as $T \rightarrow T_c$ (see details in Refs~\cite{Das:2015ana,Scardina:2017ipo}). 
Numerically we solve the RBT equation using the test particle method, where the collision integral is solved by Monte Carlo method based on stochastic interpretation of transition amplitude~\cite{Xu:2004mz,Ferini:2008he,Plumari:2012ep}. 
In our calculations the viscosity is fixed by determining the total isotropic cross section according to the Chapman-Enskog approximation:
\begin{eqnarray}
\eta=f(z)\frac{T}{\sigma}
\end{eqnarray}
where $z = m/T$ while the function $f(z)$ is defined by the following expression
\begin{eqnarray}
f(z)=\frac{15}{16}\frac{\big( z^2 K_3(z)\big)^2}{(15 z^2 +2)K_2(2z)+(3 z^3+49 z)K_3(2z)}
\end{eqnarray}
where $K_n(z)$ are the modified Bessel functions. As shown in Ref~\cite{Plumari:2012ep} the above expression for the shear viscosity $\eta$ is in quite good agreement with the Green-Kubo formula.

At the end of the evolution at freeze-out temperature, we use the hybrid hadronization approach via coalescence plus fragmentation as shown in Refs~\cite{Plumari:2017ntm}.

In our simulation for parton initial conditions, we have employed a modified Monte Carlo Glauber model, used in \cite{Sun:2019gxg} and inspired by wounded
quark model.
In this model each nucleon is composed by three constituent quarks that are randomly distributed within each nucleon according to the distribution $dN/dr=\frac{r^2}{r_0^3}e^{-r/r_0}$ with $r_0=0.3$ $fm$ and the final center of mass of the three quarks is translated to the position of the nucleon.
The positions of nucleons in $Pb$ nuclei are instead distributed according to the standard Woods–Saxon distribution.
In this way we generate the wounded quark profile using Monte-Carlo Glauber model and we decide whether each quark pair from target and projectile can collide or not with a probability
$p = e^{-\pi r^2}/\sigma_{qq}$ with $\sigma_{qq} = 13.6$ $mb$ in 5.02 ATeV $Pb + Pb$ collision \cite{Bozek:2016kpf}. Finally, the total initial parton distribution is given by:
\begin{equation}\label{profile_qu}
    \frac{dN}{d^2\textbf{x}_\perp d\eta}=\sum_{i=1}^{N_{part}}n_i\rho_\perp(\textbf{x}_\perp-\textbf{x}_i)\rho(\eta)
\end{equation}
where $N_{part}$ is the number of participant quarks while $\textbf{x}_i$ and $n_i$ are the transverse position of each participant quark and the number of partons generated by each participant respectively. The profile in the transverse position $\rho_\perp(\textbf{x}_\perp)$ is expressed as follows:
\begin{equation}\label{profile_gau}
   \rho_\perp(\textbf{x}_\perp)=\frac{1}{2\pi\sigma^2}e^{-\frac{\textbf{x}_\perp^2}{2\sigma^2}}.
\end{equation}
While the profile for the distribution of the spatial rapidity $\rho(\eta)$ is given by longitudinal boost invariance in pseudo-rapidity.
In Eq.\ref{profile_qu} the numbers $n_i$ fix the number of partons produced in each event. This number is given by $n_0 N$ where $N$ is sampled according to a negative binomial distribution given by
\begin{equation}
  P(N)=\frac{\Gamma(N+k)\bar n^N k^k}{\Gamma(k) N!(\bar n+k)^{N+k}}  
\end{equation}
where $n_0$ is fixed in order to have that the final charged particle multiplicity is same as that measured in experiments. This model is inspired by the distribution of the number of particles produced in $pp$ collisions that fluctuates according to a negative binomial distribution. 
In order to reproduce the  distribution of charged particles measured by ALICE Collaboration the parameters have been fixed to $k= 0.224$, $\bar{n} = 1.621$ and $n_0 \approx 1.88$.

In momentum space we have considered a mixture of a \textit{soft} and an \textit{hard} component, with the former consisting of quarks and gluons with an initial thermal equilibrium distribution, and the latter being equivalent to the products of initial binary pQCD collision consisting of 'minijets'. The soft partons are distributed according to the following 
\begin{eqnarray}
\label{distr_light}
    & &\frac{dN}{d^2\mathbf{x}_\perp d \eta}=\frac{g \tau_0}{(2\pi)^2} p_T m_T \times \\
    &\times &\cosh{(\eta-y)} \exp\left(-\frac{m_T\cosh{(\eta - y)}}{T(\mathbf{x}_\perp, \eta)}\right) dp_T dy
\end{eqnarray}
where $g=2\times8+3\times2\times6=52$ is the degree of freedom of partons (three flavor quarks and gluons), $\tau_0=0.4 fm/c$ is the termalization time and $m_T=\sqrt{m^2+p_T^2}$ is the transverse mass of light partons.
The transverse momentum distributions of minijets at mid-rapidity are taken from  pQCD calculations and we have used the spectra of gluons and quarks from CUJET Collaboration \cite{Xu:2015bbz} for $pp$ collisions at 5.02 TeV.

In coordinate space we initialize the charm quark distribution in accordance with the number of binary nucleon-nucleon collisions, $N_{coll}$, from the Monte Carlo Glauber model. Finally, for the charm quark initial distribution in momentum space we use the   production calculated at Fixed Order + Next-to-Leading Log (FONLL)~\cite{Cacciari:2012ny} which describes the D-meson spectra in proton-proton collisions after fragmentation. For detail we refer to our earlier work in Ref.~\cite{Scardina:2017ipo}.

In our simulation when the temperature of the QGP phase goes below the quark-hadron transition temperature, $T_c=155$ MeV, the charm quark can hadronize into D-meson. For the heavy quark hadronization process, we consider a hybrid approach of hadronization by coalescence
plus fragmentation. The fragmentation is implemented using the Peterson fragmentation function~\cite{Peterson:1982ak}: $f(z) \propto \lbrack z \lbrack 1- \frac{1}{z}- \frac{\epsilon_c}{1-z} \rbrack^2 \rbrack^{-1}$ where $z=p_D/p_c$ is the momentum fraction of the D-meson fragmented from the charm quark and
$\epsilon_c$ is a free parameter to fix so that one can describe D-meson production in proton-proton collisions \cite{Moore:2004tg}. 
For the coalescence model, it is based on a Wigner formalism. In our calculation the Wigner function is assumed to be Gaussian where the widths are fixed by the mean square charged radius of the D meson, for detail we refer to our earlier work~\cite{Plumari:2017ntm}.

\section{Heavy flavour anisotropic flows $v_n(p_T)$}
In a non-central nucleus–nucleus collision, the interaction of the QGP constituents, convert the initial eccentricity of the overlap region into a final anisotropy in momentum space, characterised in terms of the Fourier coefficients $v_n(p_T)$.\\
In our model, event-by-event fluctuations generate a profile in the transverse plane $\rho_\perp(\textbf{x}_\perp)$ that change with the event and it determines in each event the initial anisotropy in coordinate  space. This is quantified in terms of different order spatial asymmetries named $\epsilon_n$ and expressed by:
 \begin{eqnarray}
     \epsilon_n&=&\frac{r_\perp^n \cos[n(\phi-\Psi_n)]}{r_\perp^n} \\
     \Psi_n&=&\frac{1}{n} \arctan\frac{r_\perp^n \sin(n\phi)}{r_\perp^n \cos(n\phi)}
 \end{eqnarray}
where $r_\perp=\sqrt{x^2+y^2}$ and $\phi=\arctan(y/x)$. \\
The results shown in this paper have been obtained using the two particle correlation method to calculate elliptic ($v_2$) and triangular flow ($v_3$) and $v_4$ \cite{Chatrchyan:2011eka,Chatrchyan:2012wg}.
We have found that the number of test particles $N_{\rm{test}} \sim 350$ and with the lattice discretization $\Delta x=\Delta y=0.3$ fm in the simulations guarantee a convergence 
for $v_{2,3}(p_T)$ up to $p_T \sim $8 GeV$/c$.\\
\begin{figure}[h]
    \centering
    \includegraphics[width=.47\textwidth]{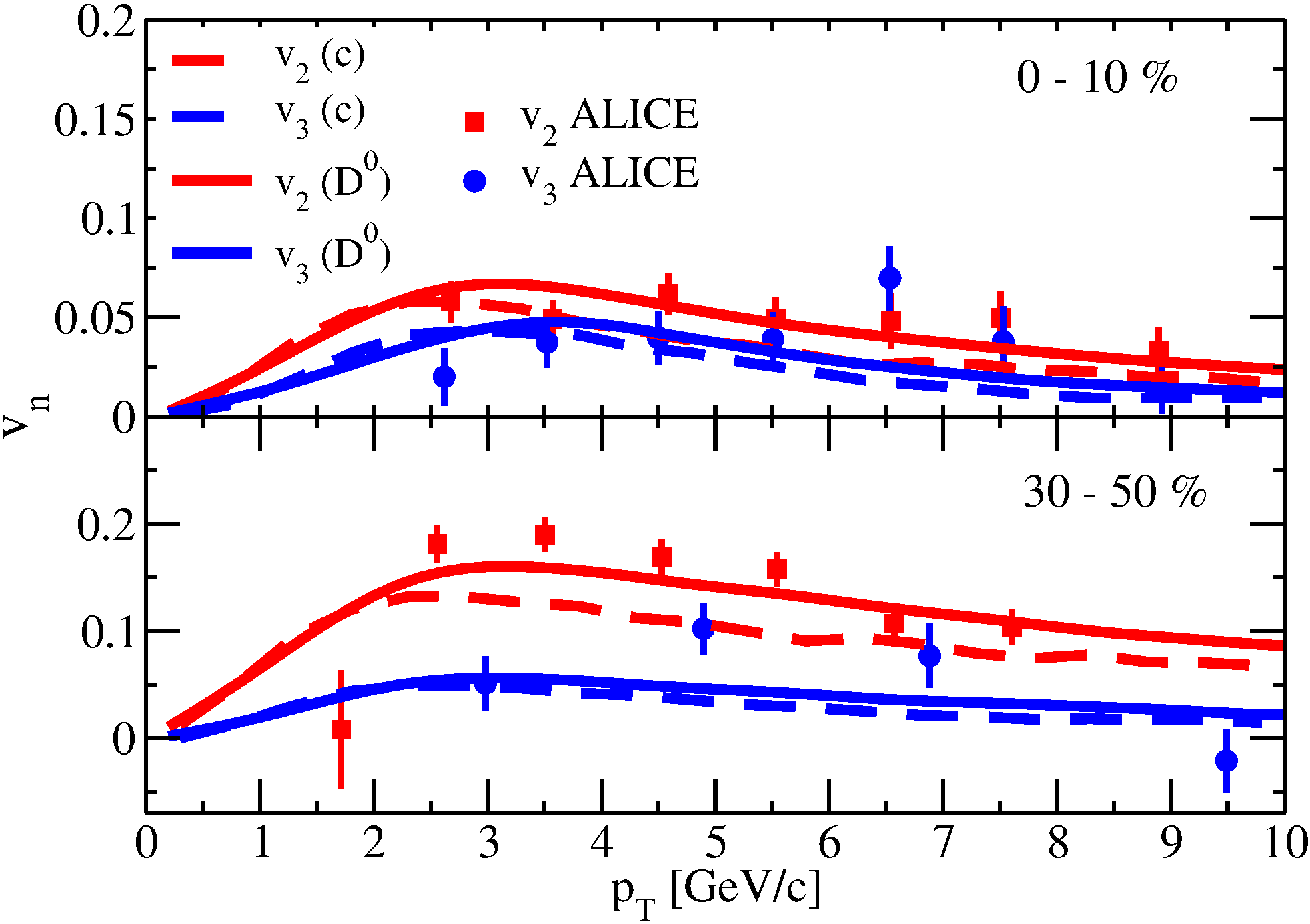}
    \caption{$v_2$ and $v_3$ of charms quarks (dashed lines) and $D$ mesons (solid lines) as a function of $p_T$  at mid-rapidity and for centrality classes $0-10\%$ (upper panel) and $30-50\%$ (lower panel) for $PbPb$ collisions at $5.02$ $ATeV$. Data are taken from ALICE measurements \cite{ALICE:2020iug}.}
    \label{fig:v2_v3}
\end{figure}
In Fig. \ref{fig:v2_v3}, we show the D mesons anisotropic flow coefficients $v_2(p_T)$ and $v_3(p_T)$ at mid-rapidity for $Pb+Pb$ collisions at $5.02 \, ATeV$ for central ($0-10\%$) in the upper panel and semi-peripheral($30-50\%$) collisions in the bottom panel. The red line corresponds to $v_2(p_T)$ and blue line to $v_3(p_T)$ both for charm quarks (dashed line) and $D$ mesons(solid line).
The charm quarks $v_2$ and $v_3$ are different from zero in both centrality classes suggesting that charm quarks take part in the collective motion. We observe that the $v_2$ is sensitive to the centrality of the collisions while the $v_3$ has a weak dependence with the centrality and it is comparable for both centralities. This suggests that the elliptic flow is mainly generated by the geometry of overlapping region and it is larger for larger centrality collision while triangular flow is mainly driven by the fluctuation of the triangularity $\epsilon_3$ of overlapping region , and similar results observed also for light quarks \cite{Plumari:2015cfa,Niemi:2012aj}. 
The hadronization mechanism plays a role for the $v_2$ to describe D meson anisotropic flow at intermediate $p_T$ range. The effect of coalescence corresponds to an increase of the collective flows at $p_T > 2$ $GeV$ 
due to the fact that $D$ mesons are formed by the recombination of a light and a heavy quarks. This effect is mainly due to the elliptic flow of light quarks which is larger than the one of charm quarks. The process of the couple recombination gives as a result a final hadron elliptic flow  larger respect to the simple anisotropy of the initial charm, because it carries information from both quarks anisotropies. On the other hand, $D$ mesons coming from fragmentation have always a smaller momentum with respect to the original charm quark, as a result the fragmentation contribution to the elliptic flow is similar to the one at charm level but shifted to lower $p_T$. As shown in Fig. \ref{fig:v2_v3}, our results are in good agreement with the ALICE experimental data \cite{ALICE:2020iug} and capture the centrality dependence observed experimentally.

\section{Event-Shape Engineering Technique}
Further investigation into the dynamics of heavy quarks in the medium can be performed with the event-shape engineering (ESE) technique \cite{Schukraft:2012ah}. This technique is based on the observation that the event-by-event variation of the anisotropic flow coefficient $v_n$ at fixed centrality is very large \cite{ALICE:2012vgf}. This technique allows for selection of events with the same centrality but different initial geometry on the basis of the magnitude of the average bulk flow $q_2$. In this section, we present the main features of the ESE selection and our results with this approach focusing on the prediction for anisotropic flows correlations $v_n-v_m$ in the heavy flavor sector.
In order to apply the ESE technique, in our calculations the events in each centrality class are divided according to the magnitude of the second-order harmonic reduced flow vector $q_2$ which is defined as:
\begin{equation}
    q_2=\frac{|\textbf{Q}_2|}{\sqrt{M}}
\end{equation}
where $\textbf{Q}_2$ is a vector built from the azimuthal angles $\phi_k$ of the considered particles and given by
\begin{equation}
  \textbf{Q}_2=\sum_{k=1}^M e^{i2\phi_k}
\end{equation}
while $M$ is the number of charged particles in the specific range of transverse momentum $p_T$ and pseudorapidity $\eta$.
\begin{figure}[h]
    \centering
    \includegraphics[width=.5\textwidth]{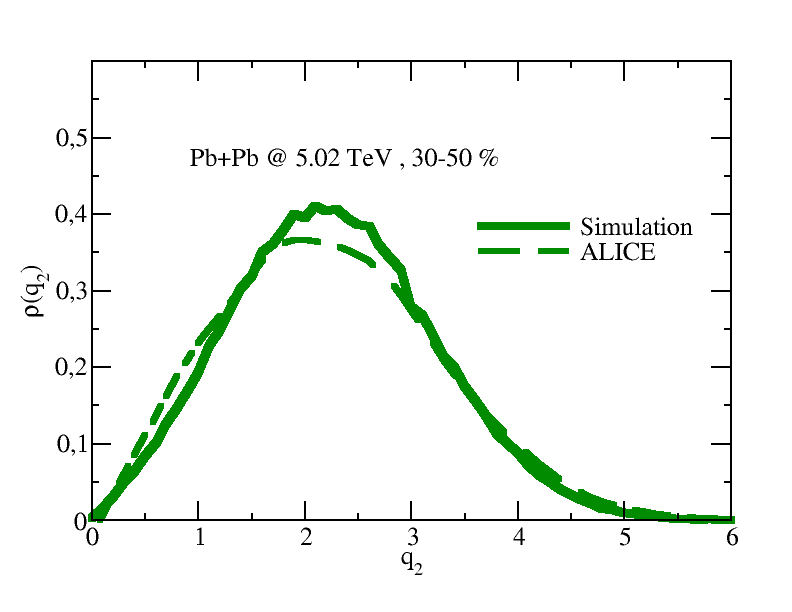}
    \caption{Distribution of charged particles at mid-rapidity ($|\eta| < 0.8$) as a function of the reduced flow vector $q_2$. In the transverse momentum range $0.2 < p_T < 5$ $GeV/c$ for the $30-50\%$ centrality class in $5.02$ $ATeV$ $PbPb$ collisions, compared to the ALICE measurements taken from \cite{ALICE:2019bdw}.}
    \label{spettroq2}
\end{figure}
In Fig. \ref{spettroq2}, it is shown $q_2$ distribution of charged particles with $|\eta| < 0.8$ and in the transverse momentum range $0.2 < p_T < 5$ $GeV/c$ for $30-50\%$ centrality class in $PbPb$ collisions at $5.02$ $ATeV$. 
The selection of the events according to the average elliptic flow was performed by defining $q_2$ percentiles in
$1\%$-wide centrality intervals to make our results comparable with the experimental data \cite{ALICE:2020iug}. Our results shown by the solid line are in in good agreement with the ALICE measurements \cite{ALICE:2019bdw} (dashed line). 
This kind of study can be performed  with the implementation of event-by-event initial state fluctuations, and we are performing one of the first calculation of this type concerning the dynamics of heavy quark.

A first study on the heavy-light flavor correlations of anisotropic flows at LHC energies with an event-by-event transport approach has been performed by means of a full Boltzmann transport approach in Ref. \cite{Plumari:2019hzp}. In this paper we extend that work and evaluate the correlations between different order anisotropic flows $v_n-v_m$ both in the light and heavy flavor sectors with the ESE technique.

\begin{figure}[h]
\includegraphics[width=.5\textwidth]{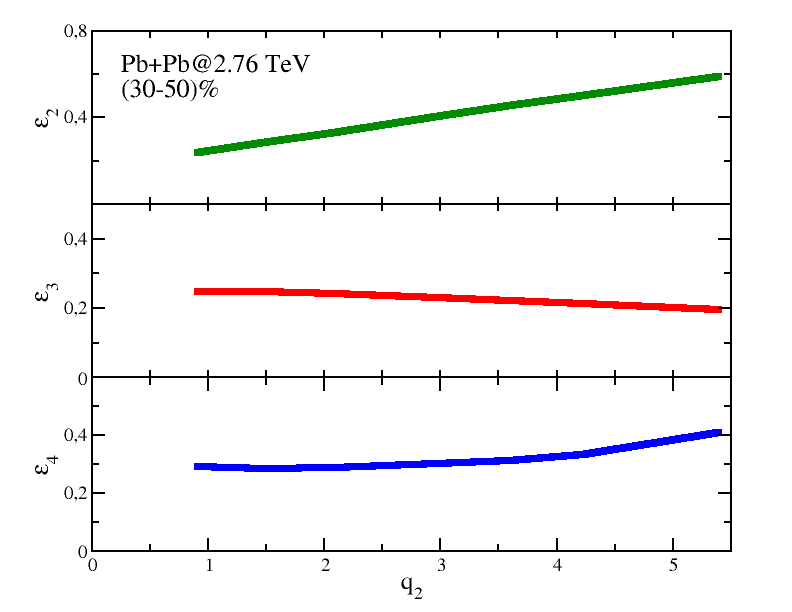}
\caption{Different order spatial eccentricities $\epsilon_{n=2,3,4}$ as a function of reduced flow vector $q_2$ for $\sqrt {s_{NN}}= 2.76$ $ATeV$ $PbPb$ collisions.}
\label{q2eccentr}
\end{figure}
In Fig. \ref{q2eccentr}, we have shown different spatial bulk eccentricities $\epsilon_n$ as a function of reduce flow vector $q_2$ for $Pb+Pb$ collisions at $\sqrt {s_{NN}}=2.76 ATeV$ in $30-50\%$ centrality class. The results show that large $q_2$ corresponds to events where the fireball has a large initial eccentricity $\epsilon_2$. The triangularity $\epsilon_3$ show a weak anti-correlation while $\epsilon_4$ receives a non linear contribution with respect to $q_2$.
The correlation between the spatial eccentricities $\epsilon_n$-$\epsilon_m$ that is present in the initial geometry gives rise to correlation between different flow harmonics $v_n$-$v_m$.

\begin{figure}
    \centering
    \includegraphics[width=0.5\textwidth]{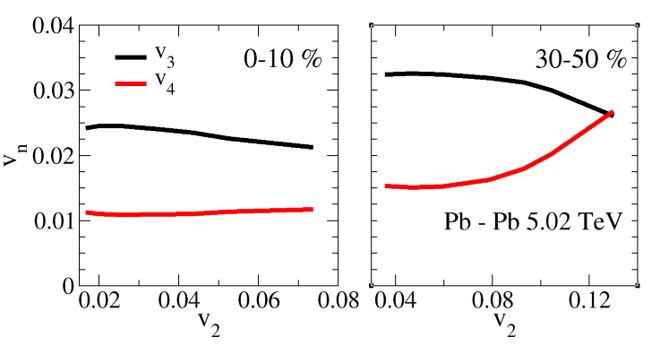}
    \caption{$v_n$ and $v_m$ ($n=2,m=3,4$) correlations for $D$ mesons in $PbPb$ collisions at $\sqrt {s_{NN}}=5.02$ $ATeV$ for both $0-10\%$ (left panel) and $30-50\%$ (right panel) centrality class.}
    \label{prediction_D}
\end{figure}

Hydrodynamic and transport calculations have shown that the response of the system to the initial spatial anisotropy is essentially linear for the second and third harmonics, meaning that the final state $v_2$ and $v_3$ are very well correlated with the second and third order eccentricities in the initial state, for small values of $\eta/s$ \cite{Heinz:2013th, Gardim:2012dc, Voloshin:2008dg, Plumari:2015cfa}.
\begin{figure}
    \centering
    \includegraphics[scale=0.8]{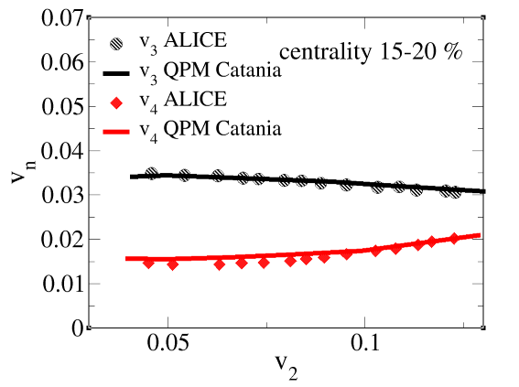}
    \caption{Correlations between $v_n$ and $v_m$ ($n=2,m=3,4$) for charged particles in $PbPb$ collisions at $\sqrt {s_{NN}}=2.76$ $ATeV$ for $15-20\%$ centrality class. The red and black points are experimental data taken from ALICE collaboration \cite{Mohapatra:2016kii}.}
    \label{charged_q2}
\end{figure}
In Fig. \ref{charged_q2} the correlations between $v_2$ and $v_3$ (solid black line), $v_2$ and $v_4$ (red solid line) for charged particles in $PbPb$ collisions at  $\sqrt {s_{NN}}=2.76 ATeV$ in $15-20\%$ centrality class are shown in comparison with the experimental data from ALICE collaboration \cite{Mohapatra:2016kii}. In these calculations, for an established centrality class, every point is determined evaluating the $v_n$ varying the $q_2$.
The trend of the $v_n - v_2$ correlations is related to the initial correlation (anti-correlation) between $\epsilon_n$ and $\epsilon_2$, because of the strong linear correlations for each harmonic $v_n \propto \epsilon_n$.
The initial anti-correlation between the $\epsilon_2$ and $\epsilon_3$ can be understood considering, for example, that 
having, for a determined centrality, a specific number of participating nucleons  can be ideally arranged to have a large $\epsilon_2$, the geometric consequence of this configuration leads to a smaller $\epsilon_3$.
These results are in agreement with other measurement performed by the ATLAS collaboration \cite{ATLAS:2015qwl}. As shown in Fig. \ref{charged_q2} the $v_4$ has a non linear correlation with the $v_2$. This behaviour can be attributed to the fact that $v_4$ receives a non-linear contribution from the initial $\epsilon_2$ that goes like $(\epsilon_2)^2$ \cite{Niemi:2012aj}. On the other hand, the $v_4$ can be parametrized as the quadrature sum of two components: one proportional to $v_2^2$ (as $v_2 \propto \epsilon_2$), and represents the non-linear component of $v_4$ driven by $(\epsilon_2)^2$, and the other contribution that in principle could be a linear component that should be independent of $v_2$ \cite{Gardim:2011xv}.
Our results for charged particles are in good agreement with the experimental data suggesting that our approach may capture the initial spatial fluctuations of the bulk matter.
In the same scheme in Fig. \ref{prediction_D} we have evaluated the  $v_n$-$v_m$ correlations in the charm sector giving predictions for $v_2$-$v_3$ and $v_2$-$v_4$ for D mesons in $PbPb$ collisions at  $\sqrt {s_{NN}}=5.02 ATeV$ in most central $0-10\%$ and semi-peripheral $30-50\%$ centrality classes. As for the light quark sector, we can clearly see an anti-correlation between $v_2$-$v_3$ and a non linear, quadratic correlation between $v_2$-$v_4$.
We observe a weak correlation between $v_3-v_2$ and $v_4-v_2$ in the central collisions and the correlation (or anti-correlation) becomes larger in more peripheral collisions. Moreover, as shown in \cite{Plumari:2019hzp} the correlation coefficient between the heavy and light quarks $v_2$  is about $C \approx 0.95$ and it remains almost flat and independent of impact parameter, suggesting a very strong correlation between the final charm quark $v_2$ and light quark $v_2$ for all the centrality investigated. On the other hand, it was approximately observed the same degree of correlation for $v_2$, $v_3$ and $v_4$. This suggest that the building up of $v_n$ of heavy quarks is driven by the $v_n$ of the bulk for $n=2,3,4$. 
This can explain why we predict a similar $v_n-v_m$ correlations of heavy quarks like the ones of the bulk. 

In order to perform a more systematic study we can now introduce a new type of observable for anisotropic flow analyses, the so-called \textit{Symmetric Cumulant} ($SC(m,n)$), proposed in \cite{Bilandzic:2013kga}. This observable is particularly useful for systems in which flow harmonics fluctuate in magnitude event-by-event, and it is defined as follows, starting from the 4-particle correlation
\begin{eqnarray}\label{SCmn}
 & & SC(m,n)= \\
 &=&<<\cos(m\phi_1+n\phi_2-m\phi_3-n\phi_4)>> \nonumber \\ & &-<<\cos(m\phi_1-m\phi_2)>><<\cos(n\phi_1-n\phi_2)>>= \nonumber \\ &=&<v_n^2v_m^2>-<v_n^2><v_m^2>
\end{eqnarray}
where the $\phi_i$ are the particle azimuthal angles, while the double averaging is a mean over all particles and over all events for a given centrality class \cite{Bilandzic:2013kga}. Notice that  for fixed values of $v_n$ and $v_m$ over all events, the symmetric cumulant as defined in Eq. \ref{SCmn}, is zero. We can get this result not only when $v_n$ and $v_m$ are fixed for all events, but also when event-by-event fluctuations of $v_n$ and $v_m$ are uncorrelated. Therefore, the symmetric cumulant is non-zero only if the event-by-event fluctuations of $v_n$ and $v_m$ are correlated. 
\begin{figure}[h]
    \centering
    \includegraphics[width=0.5\textwidth]{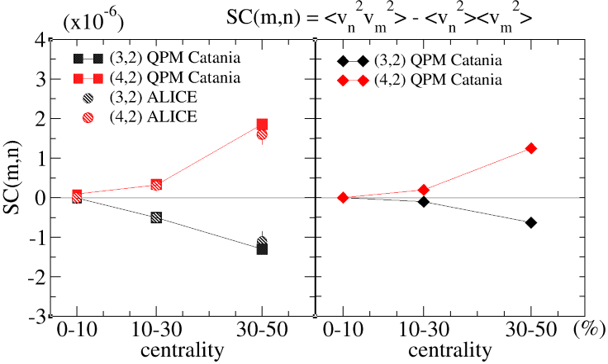}
    \caption{Symmetric Cumulant Correlator $SC(m,n)$ for both charged particles (left panel) and prediction for $D$ mesons (right panel) as a function of centrality class. In the left panel the experimental data taken from ALICE collaboration  \cite{ALICE:2021adw}}
    \label{SC}
\end{figure}
In Fig. \ref{SC}, $SC(4,2)$ (red line) and $SC(3,2)$ (black line) are shown for $PbPb$ collisions at $\sqrt {s_{NN}}=5.02 ATeV$ with $|\eta|<0.8$ and $0.5<p_T<5.0$ $GeV$ as a function of centrality. In the left panel we show the $SC(m,n)$ for charged particles that we compare with the ALICE experimental data \cite{ALICE:2021adw}, predictions for $D$ mesons are shown in the right panel. The results for charged particles are in good agreement with the experimental measurements by ALICE collaboration. The $SC(4,2)$ is positive, indicating that for a given centrality interval, events with larger $v_2$ on average have larger $v_4$. On the other hand, the $SC(3,2)$ is negative indicating an anti-correlation between $v_2$ and $v_3$ at fixed centrality. This confirm exactly what was observed within the event-shape selected $v_n$–$v_m$ correlations.
As shown in the right panel we predict a weaker but similar behaviour with the centrality for the Symmetric Cumulant Correlator $SC(m,n)$ for $D$ mesons anisotropic flows. For both cases we observe that $SC(4,2)$ and $SC(3,2)$ decrease considerably in magnitude from peripheral to central events. This large decrease 
does not imply that the correlations between these harmonics becomes negligible for centrality class smaller then $(0-10)\%$. This decreasing in the magnitude of the correlation occurs since the $SC \sim v^4_n$ and for very central events the $v_n$ (especially $v_2$) become quite small.

\begin{figure}[h]
\centering
\includegraphics[width=.45\textwidth]{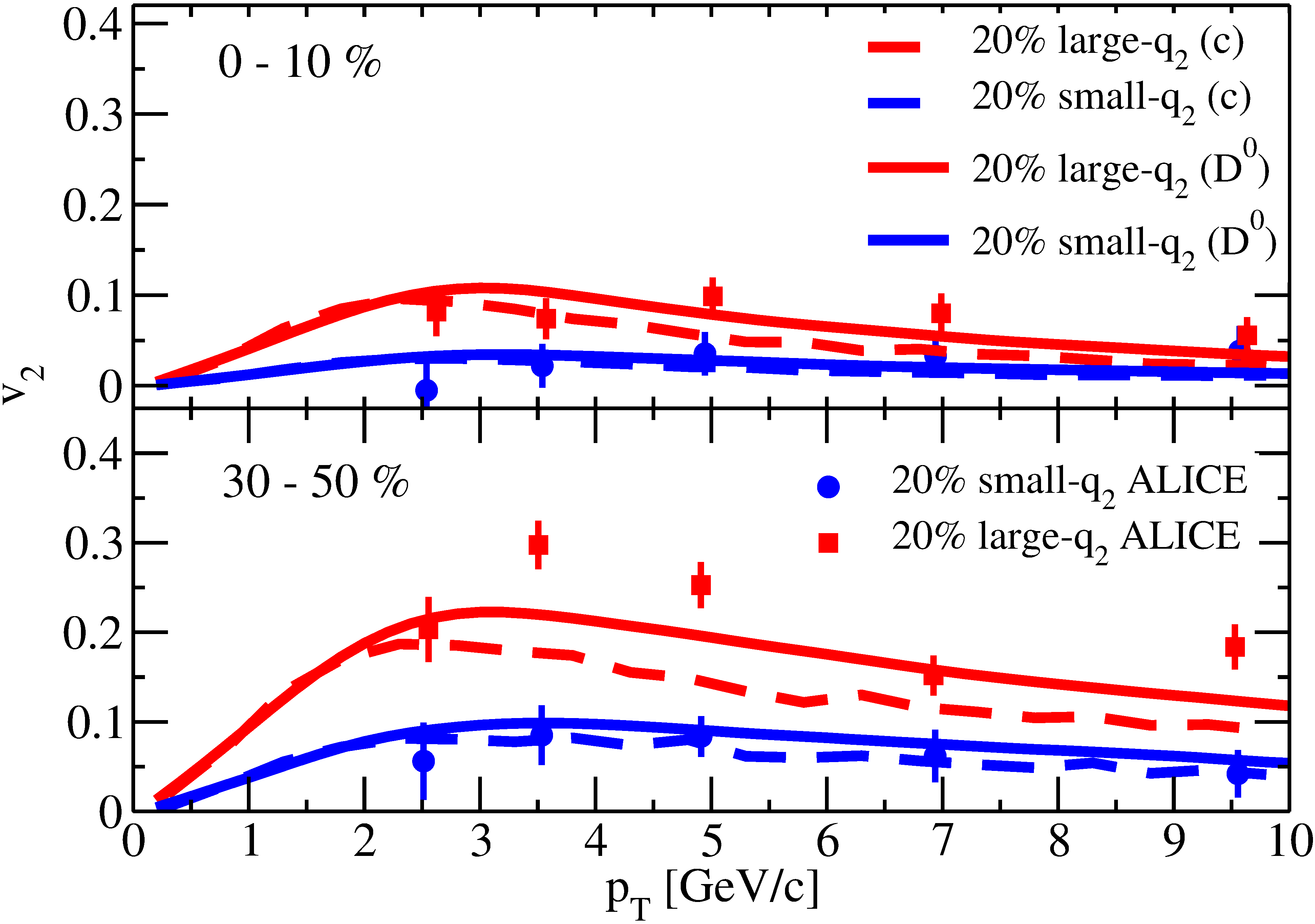}
\caption{Elliptic flow $v_2$ at mid-rapidity for $D$ mesons as a function of $p_T$ in the $20\%$ large-$q_2$ and $20\%$ small-$q_2$ events for
the centrality classes $0-10\%$ (upper panel) and $30-50\%$ (lower
panel) for $PbPb$ collisions at $5.02$ $ATeV$. Dashed lines refer to charm quarks $v_2$ while solid lines refer to D mesons $v_2$. 
The experimental data are taken from ALICE measurements \cite{ALICE:2020iug}.}
\label{v2_large_small}
\end{figure}
\begin{figure}[h!]
\centering
\includegraphics[width=.45\textwidth]{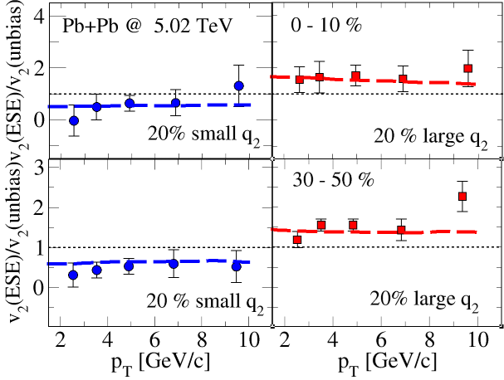}
\caption{
Ratio of the average D meson $v_2$ coefficients measured in the $20\%$ large-$q_2$ (right panel) and $20\%$ small-$q_2$ (left panel) events with respect to that of the unbiased sample as a function of $p_T$. These calculation are for for $PbPb$ collisions at $5.02$ $ATeV$ for two centralities $0-10\%$ (upper panels) and $30-50\%$ (lower panels). The experimental data are taken from ALICE measurements \cite{ALICE:2020iug}.}
\label{ratio_v2_large_small}
\end{figure}
In Fig.\ref{v2_large_small} we show the results for the $D$ mesons $v_2$ at mid-rapidity for two centralities and for different $q_2$ selections: $20\%$ large-$q_2$ (red solid and dashed lines) and $20\%$ small-$q_2$ (blue solid and dashed lines) events. As shown in Fig. \ref{v2_large_small}, the $v_2$ of $20\%$ large-$q_2$ (red solid line) events is quite larger than the $v_2$ of $20\%$ small-$q_2$ events (blue solid line) for most central $0-10\%$ and semi-peripheral $30-50\%$ centrality classes. 
This result suggests that the selection in $q_2$ distinguish between the initial eccentricities, large (small) $q_2$ means large (small) eccentricity, that consequently develop in large or small transverse flow translated into the final azimuthal anisotropy in momentum space for both bulk and D-meson. 
This difference increases with the collision centrality. Our results are compared to the ALICE measurements \cite{ALICE:2020iug}, where the good agreement shows that our model can essentially capture the strong coupling between the bulk medium and heavy quark.
Finally, in Fig. \ref{ratio_v2_large_small} we show the ratio as a function of the transverse momentum between the D meson $v_2$ in events characterized by large or small $q_2$ divided by the $v_2$ in unbiased events. We find, for these ratios, a difference of about $50\%$ between the $q_2$-selected and the unbiased events in both centrality class, with no dependence  on the transverse momentum, this behaviour might suggest that the ESE selection is related to a global property of the events. 

\section{Conclusion}
In this paper we have studied the propagation of charm quarks in the quark-gluon plasma (QGP) by means 
of an event-by-event transport approach. In particular we have solved the relativistic Boltzmann transport 
equation with an event-by-event fluctuating initial condition which allow us to access the study of high order 
anisotropic flow coefficients $v_n$ for charmed mesons.
We have studied the D meson $v_n$ at LHC energies in different centrality class selection. In particular, we have evaluated 
for the first time the heavy flavor $v_n-v_m$ correlations with the Event-Shape Engineering (ESE) technique consisting 
in selecting events in the same centrality class but characterized by different average elliptic anisotropy. 
With our approach we have a good agreement with the available experimental data from ALICE collaborations for the bulk. 
Furthermore, we have provided predictions for charmed hadrons $v_n-v_m$ correlations and we predict a weaker but similar correlations 
behaviour respect to the light flavours reflected in a smaller Symmetric Cumulant Correlator $SC(m, n)$ for D mesons.
Like for the light flavor we observe that $SC(4, 2)$ is positive, indicating that for a given centrality
interval, events with larger $v_2$ on average have larger $v_4$ while the $SC(3, 2)$ is negative indicating
an anti-correlation between $v_2$ and $v_3$ at fixed centrality. For both cases we observe that $SC(4, 2)$ and $SC(3, 2)$ 
decrease from peripheral to central events.
Finally, we have found that the selection based on the event-shape leads to a visible effect on the final particle elliptic flow. 
We have shown that the $v_2$ of $20\%$ large-$q_2$ events is larger than the $v_2$ of $20\%$ small-$q_2$ 
events for both most central $0-10\%$ and semi-peripheral $30-50\%$ centrality class, suggesting a correlation 
between the D-meson azimuthal anisotropy and the collective expansion of the bulk matter. 
We notice that the sensitivity with the event-shape selection increase with the collision centrality. Our 
results are in good agreement with the experimental data from the ALICE coll. \cite{ALICE:2020iug}.

\subsection*{Acknowledgments}
S.P. acknowledges the funding from UniCT under ‘Linea di intervento 3’ (HQsmall Grant). V.G. acknowledges the funding from UniCT under ‘Linea di intervento 2’ (HQCDyn Grant). This work was supported by the European Union's Horizon 2020 research and innovation program Strong 2020 under grant agreement No. 824093.


\begin{thebibliography}{66}
\expandafter\ifx\csname natexlab\endcsname\relax\def\natexlab#1{#1}\fi
\expandafter\ifx\csname bibnamefont\endcsname\relax
  \def\bibnamefont#1{#1}\fi
\expandafter\ifx\csname bibfnamefont\endcsname\relax
  \def\bibfnamefont#1{#1}\fi
\expandafter\ifx\csname citenamefont\endcsname\relax
  \def\citenamefont#1{#1}\fi
\expandafter\ifx\csname url\endcsname\relax
  \def\url#1{\texttt{#1}}\fi
\expandafter\ifx\csname urlprefix\endcsname\relax\def\urlprefix{URL }\fi
\providecommand{\bibinfo}[2]{#2}
\providecommand{\eprint}[2][]{\url{#2}}

\bibitem[{\citenamefont{Abelev et~al.}(2007)}]{STAR:2006btx}
\bibinfo{author}{\bibfnamefont{B.~I.} \bibnamefont{Abelev}}
  \bibnamefont{et~al.} (\bibinfo{collaboration}{STAR}), \bibinfo{journal}{Phys.
  Rev. Lett.} \textbf{\bibinfo{volume}{98}}, \bibinfo{pages}{192301}
  (\bibinfo{year}{2007}), \bibinfo{note}{[Erratum: Phys.Rev.Lett. 106, 159902
  (2011)]}, \eprint{nucl-ex/0607012}.

\bibitem[{\citenamefont{Adler et~al.}(2006)}]{PHENIX:2005nhb}
\bibinfo{author}{\bibfnamefont{S.~S.} \bibnamefont{Adler}} \bibnamefont{et~al.}
  (\bibinfo{collaboration}{PHENIX}), \bibinfo{journal}{Phys. Rev. Lett.}
  \textbf{\bibinfo{volume}{96}}, \bibinfo{pages}{032301}
  (\bibinfo{year}{2006}), \eprint{nucl-ex/0510047}.

\bibitem[{\citenamefont{Adam et~al.}(2016{\natexlab{a}})}]{ALICE:2015vxz}
\bibinfo{author}{\bibfnamefont{J.}~\bibnamefont{Adam}} \bibnamefont{et~al.}
  (\bibinfo{collaboration}{ALICE}), \bibinfo{journal}{JHEP}
  \textbf{\bibinfo{volume}{03}}, \bibinfo{pages}{081}
  (\bibinfo{year}{2016}{\natexlab{a}}), \eprint{1509.06888}.

\bibitem[{\citenamefont{Adare et~al.}(2007)}]{PHENIX:2006iih}
\bibinfo{author}{\bibfnamefont{A.}~\bibnamefont{Adare}} \bibnamefont{et~al.}
  (\bibinfo{collaboration}{PHENIX}), \bibinfo{journal}{Phys. Rev. Lett.}
  \textbf{\bibinfo{volume}{98}}, \bibinfo{pages}{172301}
  (\bibinfo{year}{2007}), \eprint{nucl-ex/0611018}.

\bibitem[{\citenamefont{Abelev et~al.}(2014)}]{ALICE:2014qvj}
\bibinfo{author}{\bibfnamefont{B.~B.} \bibnamefont{Abelev}}
  \bibnamefont{et~al.} (\bibinfo{collaboration}{ALICE}),
  \bibinfo{journal}{Phys. Rev. C} \textbf{\bibinfo{volume}{90}},
  \bibinfo{pages}{034904} (\bibinfo{year}{2014}), \eprint{1405.2001}.

\bibitem[{\citenamefont{van Hees et~al.}(2006)\citenamefont{van Hees, Greco,
  and Rapp}}]{vanHees:2005wb}
\bibinfo{author}{\bibfnamefont{H.}~\bibnamefont{van Hees}},
  \bibinfo{author}{\bibfnamefont{V.}~\bibnamefont{Greco}}, \bibnamefont{and}
  \bibinfo{author}{\bibfnamefont{R.}~\bibnamefont{Rapp}},
  \bibinfo{journal}{Phys. Rev. C} \textbf{\bibinfo{volume}{73}},
  \bibinfo{pages}{034913} (\bibinfo{year}{2006}), \eprint{nucl-th/0508055}.

\bibitem[{\citenamefont{van Hees et~al.}(2008)\citenamefont{van Hees,
  Mannarelli, Greco, and Rapp}}]{vanHees:2007me}
\bibinfo{author}{\bibfnamefont{H.}~\bibnamefont{van Hees}},
  \bibinfo{author}{\bibfnamefont{M.}~\bibnamefont{Mannarelli}},
  \bibinfo{author}{\bibfnamefont{V.}~\bibnamefont{Greco}}, \bibnamefont{and}
  \bibinfo{author}{\bibfnamefont{R.}~\bibnamefont{Rapp}},
  \bibinfo{journal}{Phys. Rev. Lett.} \textbf{\bibinfo{volume}{100}},
  \bibinfo{pages}{192301} (\bibinfo{year}{2008}), \eprint{0709.2884}.

\bibitem[{\citenamefont{Gossiaux and Aichelin}(2008)}]{Gossiaux:2008jv}
\bibinfo{author}{\bibfnamefont{P.~B.} \bibnamefont{Gossiaux}} \bibnamefont{and}
  \bibinfo{author}{\bibfnamefont{J.}~\bibnamefont{Aichelin}},
  \bibinfo{journal}{Phys. Rev. C} \textbf{\bibinfo{volume}{78}},
  \bibinfo{pages}{014904} (\bibinfo{year}{2008}), \eprint{0802.2525}.

\bibitem[{\citenamefont{Das et~al.}(2009)\citenamefont{Das, Alam, and
  Mohanty}}]{Das:2009vy}
\bibinfo{author}{\bibfnamefont{S.~K.} \bibnamefont{Das}},
  \bibinfo{author}{\bibfnamefont{J.-e.} \bibnamefont{Alam}}, \bibnamefont{and}
  \bibinfo{author}{\bibfnamefont{P.}~\bibnamefont{Mohanty}},
  \bibinfo{journal}{Phys. Rev. C} \textbf{\bibinfo{volume}{80}},
  \bibinfo{pages}{054916} (\bibinfo{year}{2009}), \eprint{0908.4194}.

\bibitem[{\citenamefont{Alberico et~al.}(2011)\citenamefont{Alberico, Beraudo,
  De~Pace, Molinari, Monteno, Nardi, and Prino}}]{Alberico:2011zy}
\bibinfo{author}{\bibfnamefont{W.~M.} \bibnamefont{Alberico}},
  \bibinfo{author}{\bibfnamefont{A.}~\bibnamefont{Beraudo}},
  \bibinfo{author}{\bibfnamefont{A.}~\bibnamefont{De~Pace}},
  \bibinfo{author}{\bibfnamefont{A.}~\bibnamefont{Molinari}},
  \bibinfo{author}{\bibfnamefont{M.}~\bibnamefont{Monteno}},
  \bibinfo{author}{\bibfnamefont{M.}~\bibnamefont{Nardi}}, \bibnamefont{and}
  \bibinfo{author}{\bibfnamefont{F.}~\bibnamefont{Prino}},
  \bibinfo{journal}{Eur. Phys. J. C} \textbf{\bibinfo{volume}{71}},
  \bibinfo{pages}{1666} (\bibinfo{year}{2011}), \eprint{1101.6008}.

\bibitem[{\citenamefont{Uphoff et~al.}(2012)\citenamefont{Uphoff, Fochler, Xu,
  and Greiner}}]{Uphoff:2012gb}
\bibinfo{author}{\bibfnamefont{J.}~\bibnamefont{Uphoff}},
  \bibinfo{author}{\bibfnamefont{O.}~\bibnamefont{Fochler}},
  \bibinfo{author}{\bibfnamefont{Z.}~\bibnamefont{Xu}}, \bibnamefont{and}
  \bibinfo{author}{\bibfnamefont{C.}~\bibnamefont{Greiner}},
  \bibinfo{journal}{Phys. Lett. B} \textbf{\bibinfo{volume}{717}},
  \bibinfo{pages}{430} (\bibinfo{year}{2012}), \eprint{1205.4945}.

\bibitem[{\citenamefont{Lang et~al.}(2016)\citenamefont{Lang, van Hees,
  Steinheimer, Inghirami, and Bleicher}}]{Lang:2012nqy}
\bibinfo{author}{\bibfnamefont{T.}~\bibnamefont{Lang}},
  \bibinfo{author}{\bibfnamefont{H.}~\bibnamefont{van Hees}},
  \bibinfo{author}{\bibfnamefont{J.}~\bibnamefont{Steinheimer}},
  \bibinfo{author}{\bibfnamefont{G.}~\bibnamefont{Inghirami}},
  \bibnamefont{and} \bibinfo{author}{\bibfnamefont{M.}~\bibnamefont{Bleicher}},
  \bibinfo{journal}{Phys. Rev. C} \textbf{\bibinfo{volume}{93}},
  \bibinfo{pages}{014901} (\bibinfo{year}{2016}), \eprint{1211.6912}.

\bibitem[{\citenamefont{Song et~al.}(2015)\citenamefont{Song, Berrehrah,
  Cabrera, Torres-Rincon, Tolos, Cassing, and Bratkovskaya}}]{Song:2015sfa}
\bibinfo{author}{\bibfnamefont{T.}~\bibnamefont{Song}},
  \bibinfo{author}{\bibfnamefont{H.}~\bibnamefont{Berrehrah}},
  \bibinfo{author}{\bibfnamefont{D.}~\bibnamefont{Cabrera}},
  \bibinfo{author}{\bibfnamefont{J.~M.} \bibnamefont{Torres-Rincon}},
  \bibinfo{author}{\bibfnamefont{L.}~\bibnamefont{Tolos}},
  \bibinfo{author}{\bibfnamefont{W.}~\bibnamefont{Cassing}}, \bibnamefont{and}
  \bibinfo{author}{\bibfnamefont{E.}~\bibnamefont{Bratkovskaya}},
  \bibinfo{journal}{Phys. Rev. C} \textbf{\bibinfo{volume}{92}},
  \bibinfo{pages}{014910} (\bibinfo{year}{2015}), \eprint{1503.03039}.

\bibitem[{\citenamefont{Song et~al.}(2016)\citenamefont{Song, Berrehrah,
  Cabrera, Cassing, and Bratkovskaya}}]{Song:2015ykw}
\bibinfo{author}{\bibfnamefont{T.}~\bibnamefont{Song}},
  \bibinfo{author}{\bibfnamefont{H.}~\bibnamefont{Berrehrah}},
  \bibinfo{author}{\bibfnamefont{D.}~\bibnamefont{Cabrera}},
  \bibinfo{author}{\bibfnamefont{W.}~\bibnamefont{Cassing}}, \bibnamefont{and}
  \bibinfo{author}{\bibfnamefont{E.}~\bibnamefont{Bratkovskaya}},
  \bibinfo{journal}{Phys. Rev. C} \textbf{\bibinfo{volume}{93}},
  \bibinfo{pages}{034906} (\bibinfo{year}{2016}), \eprint{1512.00891}.

\bibitem[{\citenamefont{Das et~al.}(2014)\citenamefont{Das, Scardina, Plumari,
  and Greco}}]{Das:2013kea}
\bibinfo{author}{\bibfnamefont{S.~K.} \bibnamefont{Das}},
  \bibinfo{author}{\bibfnamefont{F.}~\bibnamefont{Scardina}},
  \bibinfo{author}{\bibfnamefont{S.}~\bibnamefont{Plumari}}, \bibnamefont{and}
  \bibinfo{author}{\bibfnamefont{V.}~\bibnamefont{Greco}},
  \bibinfo{journal}{Phys. Rev. C} \textbf{\bibinfo{volume}{90}},
  \bibinfo{pages}{044901} (\bibinfo{year}{2014}), \eprint{1312.6857}.

\bibitem[{\citenamefont{Cao et~al.}(2015)\citenamefont{Cao, Qin, and
  Bass}}]{Cao:2015hia}
\bibinfo{author}{\bibfnamefont{S.}~\bibnamefont{Cao}},
  \bibinfo{author}{\bibfnamefont{G.-Y.} \bibnamefont{Qin}}, \bibnamefont{and}
  \bibinfo{author}{\bibfnamefont{S.~A.} \bibnamefont{Bass}},
  \bibinfo{journal}{Phys. Rev. C} \textbf{\bibinfo{volume}{92}},
  \bibinfo{pages}{024907} (\bibinfo{year}{2015}), \eprint{1505.01413}.

\bibitem[{\citenamefont{Das et~al.}(2015)\citenamefont{Das, Scardina, Plumari,
  and Greco}}]{Das:2015ana}
\bibinfo{author}{\bibfnamefont{S.~K.} \bibnamefont{Das}},
  \bibinfo{author}{\bibfnamefont{F.}~\bibnamefont{Scardina}},
  \bibinfo{author}{\bibfnamefont{S.}~\bibnamefont{Plumari}}, \bibnamefont{and}
  \bibinfo{author}{\bibfnamefont{V.}~\bibnamefont{Greco}},
  \bibinfo{journal}{Phys. Lett. B} \textbf{\bibinfo{volume}{747}},
  \bibinfo{pages}{260} (\bibinfo{year}{2015}), \eprint{1502.03757}.

\bibitem[{\citenamefont{Cao et~al.}(2018)\citenamefont{Cao, Luo, Qin, and
  Wang}}]{Cao:2017hhk}
\bibinfo{author}{\bibfnamefont{S.}~\bibnamefont{Cao}},
  \bibinfo{author}{\bibfnamefont{T.}~\bibnamefont{Luo}},
  \bibinfo{author}{\bibfnamefont{G.-Y.} \bibnamefont{Qin}}, \bibnamefont{and}
  \bibinfo{author}{\bibfnamefont{X.-N.} \bibnamefont{Wang}},
  \bibinfo{journal}{Phys. Lett. B} \textbf{\bibinfo{volume}{777}},
  \bibinfo{pages}{255} (\bibinfo{year}{2018}), \eprint{1703.00822}.

\bibitem[{\citenamefont{Das et~al.}(2017)\citenamefont{Das, Ruggieri, Scardina,
  Plumari, and Greco}}]{Das:2017dsh}
\bibinfo{author}{\bibfnamefont{S.~K.} \bibnamefont{Das}},
  \bibinfo{author}{\bibfnamefont{M.}~\bibnamefont{Ruggieri}},
  \bibinfo{author}{\bibfnamefont{F.}~\bibnamefont{Scardina}},
  \bibinfo{author}{\bibfnamefont{S.}~\bibnamefont{Plumari}}, \bibnamefont{and}
  \bibinfo{author}{\bibfnamefont{V.}~\bibnamefont{Greco}}, \bibinfo{journal}{J.
  Phys. G} \textbf{\bibinfo{volume}{44}}, \bibinfo{pages}{095102}
  (\bibinfo{year}{2017}), \eprint{1701.05123}.

\bibitem[{\citenamefont{Sun et~al.}(2019)\citenamefont{Sun, Coci, Das, Plumari,
  Ruggieri, and Greco}}]{Sun:2019fud}
\bibinfo{author}{\bibfnamefont{Y.}~\bibnamefont{Sun}},
  \bibinfo{author}{\bibfnamefont{G.}~\bibnamefont{Coci}},
  \bibinfo{author}{\bibfnamefont{S.~K.} \bibnamefont{Das}},
  \bibinfo{author}{\bibfnamefont{S.}~\bibnamefont{Plumari}},
  \bibinfo{author}{\bibfnamefont{M.}~\bibnamefont{Ruggieri}}, \bibnamefont{and}
  \bibinfo{author}{\bibfnamefont{V.}~\bibnamefont{Greco}},
  \bibinfo{journal}{Phys. Lett. B} \textbf{\bibinfo{volume}{798}},
  \bibinfo{pages}{134933} (\bibinfo{year}{2019}), \eprint{1902.06254}.

\bibitem[{\citenamefont{Cao et~al.}(2019)}]{Cao:2018ews}
\bibinfo{author}{\bibfnamefont{S.}~\bibnamefont{Cao}} \bibnamefont{et~al.},
  \bibinfo{journal}{Phys. Rev. C} \textbf{\bibinfo{volume}{99}},
  \bibinfo{pages}{054907} (\bibinfo{year}{2019}), \eprint{1809.07894}.

\bibitem[{\citenamefont{Beraudo et~al.}(2018)}]{Rapp:2018qla}
\bibinfo{author}{\bibfnamefont{A.}~\bibnamefont{Beraudo}} \bibnamefont{et~al.},
  \bibinfo{journal}{Nucl. Phys. A} \textbf{\bibinfo{volume}{979}},
  \bibinfo{pages}{21} (\bibinfo{year}{2018}), \eprint{1803.03824}.

\bibitem[{\citenamefont{Plumari et~al.}(2018)\citenamefont{Plumari, Minissale,
  Das, Coci, and Greco}}]{Plumari:2017ntm}
\bibinfo{author}{\bibfnamefont{S.}~\bibnamefont{Plumari}},
  \bibinfo{author}{\bibfnamefont{V.}~\bibnamefont{Minissale}},
  \bibinfo{author}{\bibfnamefont{S.~K.} \bibnamefont{Das}},
  \bibinfo{author}{\bibfnamefont{G.}~\bibnamefont{Coci}}, \bibnamefont{and}
  \bibinfo{author}{\bibfnamefont{V.}~\bibnamefont{Greco}},
  \bibinfo{journal}{Eur. Phys. J. C} \textbf{\bibinfo{volume}{78}},
  \bibinfo{pages}{348} (\bibinfo{year}{2018}).

\bibitem[{\citenamefont{Minissale et~al.}(2021)\citenamefont{Minissale,
  Plumari, and Greco}}]{Minissale:2020bif}
\bibinfo{author}{\bibfnamefont{V.}~\bibnamefont{Minissale}},
  \bibinfo{author}{\bibfnamefont{S.}~\bibnamefont{Plumari}}, \bibnamefont{and}
  \bibinfo{author}{\bibfnamefont{V.}~\bibnamefont{Greco}},
  \bibinfo{journal}{Phys. Lett. B} \textbf{\bibinfo{volume}{821}},
  \bibinfo{pages}{136622} (\bibinfo{year}{2021}), \eprint{2012.12001}.

\bibitem[{\citenamefont{Nahrgang et~al.}(2015)\citenamefont{Nahrgang, Aichelin,
  Bass, Gossiaux, and Werner}}]{Nahrgang:2014vza}
\bibinfo{author}{\bibfnamefont{M.}~\bibnamefont{Nahrgang}},
  \bibinfo{author}{\bibfnamefont{J.}~\bibnamefont{Aichelin}},
  \bibinfo{author}{\bibfnamefont{S.}~\bibnamefont{Bass}},
  \bibinfo{author}{\bibfnamefont{P.~B.} \bibnamefont{Gossiaux}},
  \bibnamefont{and} \bibinfo{author}{\bibfnamefont{K.}~\bibnamefont{Werner}},
  \bibinfo{journal}{Phys. Rev. C} \textbf{\bibinfo{volume}{91}},
  \bibinfo{pages}{014904} (\bibinfo{year}{2015}), \eprint{1410.5396}.

\bibitem[{\citenamefont{Prado et~al.}(2017)\citenamefont{Prado,
  Noronha-Hostler, Katz, Suaide, Noronha, Munhoz, and
  Cosentino}}]{Prado:2016szr}
\bibinfo{author}{\bibfnamefont{C.~A.~G.} \bibnamefont{Prado}},
  \bibinfo{author}{\bibfnamefont{J.}~\bibnamefont{Noronha-Hostler}},
  \bibinfo{author}{\bibfnamefont{R.}~\bibnamefont{Katz}},
  \bibinfo{author}{\bibfnamefont{A.~A.~P.} \bibnamefont{Suaide}},
  \bibinfo{author}{\bibfnamefont{J.}~\bibnamefont{Noronha}},
  \bibinfo{author}{\bibfnamefont{M.~G.} \bibnamefont{Munhoz}},
  \bibnamefont{and} \bibinfo{author}{\bibfnamefont{M.~R.}
  \bibnamefont{Cosentino}}, \bibinfo{journal}{Phys. Rev. C}
  \textbf{\bibinfo{volume}{96}}, \bibinfo{pages}{064903}
  (\bibinfo{year}{2017}), \eprint{1611.02965}.

\bibitem[{\citenamefont{Beraudo et~al.}(2019)\citenamefont{Beraudo, De~Pace,
  Monteno, Nardi, and Prino}}]{Beraudo:2018tpr}
\bibinfo{author}{\bibfnamefont{A.}~\bibnamefont{Beraudo}},
  \bibinfo{author}{\bibfnamefont{A.}~\bibnamefont{De~Pace}},
  \bibinfo{author}{\bibfnamefont{M.}~\bibnamefont{Monteno}},
  \bibinfo{author}{\bibfnamefont{M.}~\bibnamefont{Nardi}}, \bibnamefont{and}
  \bibinfo{author}{\bibfnamefont{F.}~\bibnamefont{Prino}},
  \bibinfo{journal}{Eur. Phys. J. C} \textbf{\bibinfo{volume}{79}},
  \bibinfo{pages}{494} (\bibinfo{year}{2019}), \eprint{1812.08337}.

\bibitem[{\citenamefont{Katz et~al.}(2020)\citenamefont{Katz, Prado,
  Noronha-Hostler, Noronha, and Suaide}}]{Katz:2019fkc}
\bibinfo{author}{\bibfnamefont{R.}~\bibnamefont{Katz}},
  \bibinfo{author}{\bibfnamefont{C.~A.~G.} \bibnamefont{Prado}},
  \bibinfo{author}{\bibfnamefont{J.}~\bibnamefont{Noronha-Hostler}},
  \bibinfo{author}{\bibfnamefont{J.}~\bibnamefont{Noronha}}, \bibnamefont{and}
  \bibinfo{author}{\bibfnamefont{A.~A.~P.} \bibnamefont{Suaide}},
  \bibinfo{journal}{Phys. Rev. C} \textbf{\bibinfo{volume}{102}},
  \bibinfo{pages}{024906} (\bibinfo{year}{2020}), \eprint{1906.10768}.

\bibitem[{\citenamefont{Plumari et~al.}(2020)\citenamefont{Plumari, Coci,
  Minissale, Das, Sun, and Greco}}]{Plumari:2019hzp}
\bibinfo{author}{\bibfnamefont{S.}~\bibnamefont{Plumari}},
  \bibinfo{author}{\bibfnamefont{G.}~\bibnamefont{Coci}},
  \bibinfo{author}{\bibfnamefont{V.}~\bibnamefont{Minissale}},
  \bibinfo{author}{\bibfnamefont{S.~K.} \bibnamefont{Das}},
  \bibinfo{author}{\bibfnamefont{Y.}~\bibnamefont{Sun}}, \bibnamefont{and}
  \bibinfo{author}{\bibfnamefont{V.}~\bibnamefont{Greco}},
  \bibinfo{journal}{Phys. Lett. B} \textbf{\bibinfo{volume}{805}},
  \bibinfo{pages}{135460} (\bibinfo{year}{2020}), \eprint{1912.09350}.

\bibitem[{\citenamefont{Schukraft et~al.}(2013)\citenamefont{Schukraft,
  Timmins, and Voloshin}}]{Schukraft:2012ah}
\bibinfo{author}{\bibfnamefont{J.}~\bibnamefont{Schukraft}},
  \bibinfo{author}{\bibfnamefont{A.}~\bibnamefont{Timmins}}, \bibnamefont{and}
  \bibinfo{author}{\bibfnamefont{S.~A.} \bibnamefont{Voloshin}},
  \bibinfo{journal}{Phys. Lett. B} \textbf{\bibinfo{volume}{719}},
  \bibinfo{pages}{394} (\bibinfo{year}{2013}), \eprint{1208.4563}.

\bibitem[{\citenamefont{Adam et~al.}(2016{\natexlab{b}})}]{ALICE:2015lib}
\bibinfo{author}{\bibfnamefont{J.}~\bibnamefont{Adam}} \bibnamefont{et~al.}
  (\bibinfo{collaboration}{ALICE}), \bibinfo{journal}{Phys. Rev. C}
  \textbf{\bibinfo{volume}{93}}, \bibinfo{pages}{034916}
  (\bibinfo{year}{2016}{\natexlab{b}}), \eprint{1507.06194}.

\bibitem[{\citenamefont{Acharya et~al.}(2019{\natexlab{a}})}]{ALICE:2018gif}
\bibinfo{author}{\bibfnamefont{S.}~\bibnamefont{Acharya}} \bibnamefont{et~al.}
  (\bibinfo{collaboration}{ALICE}), \bibinfo{journal}{JHEP}
  \textbf{\bibinfo{volume}{02}}, \bibinfo{pages}{150}
  (\bibinfo{year}{2019}{\natexlab{a}}), \eprint{1809.09371}.

\bibitem[{\citenamefont{Plumari et~al.}(2012)\citenamefont{Plumari, Puglisi,
  Scardina, and Greco}}]{Plumari:2012ep}
\bibinfo{author}{\bibfnamefont{S.}~\bibnamefont{Plumari}},
  \bibinfo{author}{\bibfnamefont{A.}~\bibnamefont{Puglisi}},
  \bibinfo{author}{\bibfnamefont{F.}~\bibnamefont{Scardina}}, \bibnamefont{and}
  \bibinfo{author}{\bibfnamefont{V.}~\bibnamefont{Greco}},
  \bibinfo{journal}{Phys. Rev. C} \textbf{\bibinfo{volume}{86}},
  \bibinfo{pages}{054902} (\bibinfo{year}{2012}), \eprint{1208.0481}.

\bibitem[{\citenamefont{Ruggieri et~al.}(2013)\citenamefont{Ruggieri, Scardina,
  Plumari, and Greco}}]{Ruggieri:2013bda}
\bibinfo{author}{\bibfnamefont{M.}~\bibnamefont{Ruggieri}},
  \bibinfo{author}{\bibfnamefont{F.}~\bibnamefont{Scardina}},
  \bibinfo{author}{\bibfnamefont{S.}~\bibnamefont{Plumari}}, \bibnamefont{and}
  \bibinfo{author}{\bibfnamefont{V.}~\bibnamefont{Greco}},
  \bibinfo{journal}{Phys. Lett.} \textbf{\bibinfo{volume}{B727}},
  \bibinfo{pages}{177} (\bibinfo{year}{2013}).

\bibitem[{\citenamefont{Scardina et~al.}(2013)\citenamefont{Scardina, Colonna,
  Plumari, and Greco}}]{Scardina:2012mik}
\bibinfo{author}{\bibfnamefont{F.}~\bibnamefont{Scardina}},
  \bibinfo{author}{\bibfnamefont{M.}~\bibnamefont{Colonna}},
  \bibinfo{author}{\bibfnamefont{S.}~\bibnamefont{Plumari}}, \bibnamefont{and}
  \bibinfo{author}{\bibfnamefont{V.}~\bibnamefont{Greco}},
  \bibinfo{journal}{Phys. Lett. B} \textbf{\bibinfo{volume}{724}},
  \bibinfo{pages}{296} (\bibinfo{year}{2013}), \eprint{1202.2262}.

\bibitem[{\citenamefont{Ruggieri et~al.}(2014)\citenamefont{Ruggieri, Scardina,
  Plumari, and Greco}}]{Ruggieri:2013ova}
\bibinfo{author}{\bibfnamefont{M.}~\bibnamefont{Ruggieri}},
  \bibinfo{author}{\bibfnamefont{F.}~\bibnamefont{Scardina}},
  \bibinfo{author}{\bibfnamefont{S.}~\bibnamefont{Plumari}}, \bibnamefont{and}
  \bibinfo{author}{\bibfnamefont{V.}~\bibnamefont{Greco}},
  \bibinfo{journal}{Phys. Rev. C} \textbf{\bibinfo{volume}{89}},
  \bibinfo{pages}{054914} (\bibinfo{year}{2014}).

\bibitem[{\citenamefont{Puglisi et~al.}(2014)\citenamefont{Puglisi, Plumari,
  and Greco}}]{Puglisi:2014sha}
\bibinfo{author}{\bibfnamefont{A.}~\bibnamefont{Puglisi}},
  \bibinfo{author}{\bibfnamefont{S.}~\bibnamefont{Plumari}}, \bibnamefont{and}
  \bibinfo{author}{\bibfnamefont{V.}~\bibnamefont{Greco}},
  \bibinfo{journal}{Phys. Rev. D} \textbf{\bibinfo{volume}{90}},
  \bibinfo{pages}{114009} (\bibinfo{year}{2014}), \eprint{1408.7043}.

\bibitem[{\citenamefont{Scardina et~al.}(2014)\citenamefont{Scardina,
  Perricone, Plumari, Ruggieri, and Greco}}]{Scardina:2014gxa}
\bibinfo{author}{\bibfnamefont{F.}~\bibnamefont{Scardina}},
  \bibinfo{author}{\bibfnamefont{D.}~\bibnamefont{Perricone}},
  \bibinfo{author}{\bibfnamefont{S.}~\bibnamefont{Plumari}},
  \bibinfo{author}{\bibfnamefont{M.}~\bibnamefont{Ruggieri}}, \bibnamefont{and}
  \bibinfo{author}{\bibfnamefont{V.}~\bibnamefont{Greco}},
  \bibinfo{journal}{Phys. Rev. C} \textbf{\bibinfo{volume}{90}},
  \bibinfo{pages}{054904} (\bibinfo{year}{2014}), \eprint{1408.1313}.

\bibitem[{\citenamefont{Plumari
  et~al.}(2015{\natexlab{a}})\citenamefont{Plumari, Guardo, Greco, and
  Ollitrault}}]{Plumari:2015sia}
\bibinfo{author}{\bibfnamefont{S.}~\bibnamefont{Plumari}},
  \bibinfo{author}{\bibfnamefont{G.~L.} \bibnamefont{Guardo}},
  \bibinfo{author}{\bibfnamefont{V.}~\bibnamefont{Greco}}, \bibnamefont{and}
  \bibinfo{author}{\bibfnamefont{J.-Y.} \bibnamefont{Ollitrault}},
  \bibinfo{journal}{Nucl. Phys.} \textbf{\bibinfo{volume}{A941}},
  \bibinfo{pages}{87} (\bibinfo{year}{2015}{\natexlab{a}}).

\bibitem[{\citenamefont{Plumari
  et~al.}(2015{\natexlab{b}})\citenamefont{Plumari, Guardo, Scardina, and
  Greco}}]{Plumari:2015cfa}
\bibinfo{author}{\bibfnamefont{S.}~\bibnamefont{Plumari}},
  \bibinfo{author}{\bibfnamefont{G.~L.} \bibnamefont{Guardo}},
  \bibinfo{author}{\bibfnamefont{F.}~\bibnamefont{Scardina}}, \bibnamefont{and}
  \bibinfo{author}{\bibfnamefont{V.}~\bibnamefont{Greco}},
  \bibinfo{journal}{Phys. Rev. C} \textbf{\bibinfo{volume}{92}},
  \bibinfo{pages}{054902} (\bibinfo{year}{2015}{\natexlab{b}}),
  \eprint{1507.05540}.

\bibitem[{\citenamefont{Scardina et~al.}(2017)\citenamefont{Scardina, Das,
  Minissale, Plumari, and Greco}}]{Scardina:2017ipo}
\bibinfo{author}{\bibfnamefont{F.}~\bibnamefont{Scardina}},
  \bibinfo{author}{\bibfnamefont{S.~K.} \bibnamefont{Das}},
  \bibinfo{author}{\bibfnamefont{V.}~\bibnamefont{Minissale}},
  \bibinfo{author}{\bibfnamefont{S.}~\bibnamefont{Plumari}}, \bibnamefont{and}
  \bibinfo{author}{\bibfnamefont{V.}~\bibnamefont{Greco}},
  \bibinfo{journal}{Phys.\ Rev.\ C} \textbf{\bibinfo{volume}{96}},
  \bibinfo{pages}{044905} (\bibinfo{year}{2017}).

\bibitem[{\citenamefont{Plumari}(2019)}]{Plumari:2019gwq}
\bibinfo{author}{\bibfnamefont{S.}~\bibnamefont{Plumari}},
  \bibinfo{journal}{Eur. Phys. J.} \textbf{\bibinfo{volume}{C79}},
  \bibinfo{pages}{2} (\bibinfo{year}{2019}).

\bibitem[{\citenamefont{Sun et~al.}(2020)\citenamefont{Sun, Plumari, and
  Greco}}]{Sun:2019gxg}
\bibinfo{author}{\bibfnamefont{Y.}~\bibnamefont{Sun}},
  \bibinfo{author}{\bibfnamefont{S.}~\bibnamefont{Plumari}}, \bibnamefont{and}
  \bibinfo{author}{\bibfnamefont{V.}~\bibnamefont{Greco}},
  \bibinfo{journal}{Eur.\ Phys.\ J.\ C} \textbf{\bibinfo{volume}{80}},
  \bibinfo{pages}{16} (\bibinfo{year}{2020}).

\bibitem[{\citenamefont{Plumari et~al.}(2013)\citenamefont{Plumari, Puglisi,
  Colonna, Scardina, and Greco}}]{Plumari:2012xz}
\bibinfo{author}{\bibfnamefont{S.}~\bibnamefont{Plumari}},
  \bibinfo{author}{\bibfnamefont{A.}~\bibnamefont{Puglisi}},
  \bibinfo{author}{\bibfnamefont{M.}~\bibnamefont{Colonna}},
  \bibinfo{author}{\bibfnamefont{F.}~\bibnamefont{Scardina}}, \bibnamefont{and}
  \bibinfo{author}{\bibfnamefont{V.}~\bibnamefont{Greco}}, \bibinfo{journal}{J.
  Phys. Conf. Ser.} \textbf{\bibinfo{volume}{420}}, \bibinfo{pages}{012029}
  (\bibinfo{year}{2013}), \eprint{1209.0601}.

\bibitem[{\citenamefont{Borsanyi et~al.}(2010)\citenamefont{Borsanyi, Endrodi,
  Fodor, Jakovac, Katz, Krieg, Ratti, and Szabo}}]{Borsanyi:2010cj}
\bibinfo{author}{\bibfnamefont{S.}~\bibnamefont{Borsanyi}},
  \bibinfo{author}{\bibfnamefont{G.}~\bibnamefont{Endrodi}},
  \bibinfo{author}{\bibfnamefont{Z.}~\bibnamefont{Fodor}},
  \bibinfo{author}{\bibfnamefont{A.}~\bibnamefont{Jakovac}},
  \bibinfo{author}{\bibfnamefont{S.~D.} \bibnamefont{Katz}},
  \bibinfo{author}{\bibfnamefont{S.}~\bibnamefont{Krieg}},
  \bibinfo{author}{\bibfnamefont{C.}~\bibnamefont{Ratti}}, \bibnamefont{and}
  \bibinfo{author}{\bibfnamefont{K.~K.} \bibnamefont{Szabo}},
  \bibinfo{journal}{JHEP} \textbf{\bibinfo{volume}{11}}, \bibinfo{pages}{077}
  (\bibinfo{year}{2010}).

\bibitem[{\citenamefont{Xu and Greiner}(2005)}]{Xu:2004mz}
\bibinfo{author}{\bibfnamefont{Z.}~\bibnamefont{Xu}} \bibnamefont{and}
  \bibinfo{author}{\bibfnamefont{C.}~\bibnamefont{Greiner}},
  \bibinfo{journal}{Phys. Rev.} \textbf{\bibinfo{volume}{C71}},
  \bibinfo{pages}{064901} (\bibinfo{year}{2005}).

\bibitem[{\citenamefont{Ferini et~al.}(2009)\citenamefont{Ferini, Colonna,
  Di~Toro, and Greco}}]{Ferini:2008he}
\bibinfo{author}{\bibfnamefont{G.}~\bibnamefont{Ferini}},
  \bibinfo{author}{\bibfnamefont{M.}~\bibnamefont{Colonna}},
  \bibinfo{author}{\bibfnamefont{M.}~\bibnamefont{Di~Toro}}, \bibnamefont{and}
  \bibinfo{author}{\bibfnamefont{V.}~\bibnamefont{Greco}},
  \bibinfo{journal}{Phys. Lett.} \textbf{\bibinfo{volume}{B670}},
  \bibinfo{pages}{325} (\bibinfo{year}{2009}).

\bibitem[{\citenamefont{Bozek et~al.}(2016)\citenamefont{Bozek, Broniowski, and
  Rybczynski}}]{Bozek:2016kpf}
\bibinfo{author}{\bibfnamefont{P.}~\bibnamefont{Bozek}},
  \bibinfo{author}{\bibfnamefont{W.}~\bibnamefont{Broniowski}},
  \bibnamefont{and}
  \bibinfo{author}{\bibfnamefont{M.}~\bibnamefont{Rybczynski}},
  \bibinfo{journal}{Phys. Rev.} \textbf{\bibinfo{volume}{C94}},
  \bibinfo{pages}{014902} (\bibinfo{year}{2016}).

\bibitem[{\citenamefont{Xu et~al.}(2016)\citenamefont{Xu, Liao, and
  Gyulassy}}]{Xu:2015bbz}
\bibinfo{author}{\bibfnamefont{J.}~\bibnamefont{Xu}},
  \bibinfo{author}{\bibfnamefont{J.}~\bibnamefont{Liao}}, \bibnamefont{and}
  \bibinfo{author}{\bibfnamefont{M.}~\bibnamefont{Gyulassy}},
  \bibinfo{journal}{JHEP} \textbf{\bibinfo{volume}{02}}, \bibinfo{pages}{169}
  (\bibinfo{year}{2016}).

\bibitem[{\citenamefont{Cacciari et~al.}(2012)\citenamefont{Cacciari, Frixione,
  Houdeau, Mangano, Nason, and Ridolfi}}]{Cacciari:2012ny}
\bibinfo{author}{\bibfnamefont{M.}~\bibnamefont{Cacciari}},
  \bibinfo{author}{\bibfnamefont{S.}~\bibnamefont{Frixione}},
  \bibinfo{author}{\bibfnamefont{N.}~\bibnamefont{Houdeau}},
  \bibinfo{author}{\bibfnamefont{M.~L.} \bibnamefont{Mangano}},
  \bibinfo{author}{\bibfnamefont{P.}~\bibnamefont{Nason}}, \bibnamefont{and}
  \bibinfo{author}{\bibfnamefont{G.}~\bibnamefont{Ridolfi}},
  \bibinfo{journal}{JHEP} \textbf{\bibinfo{volume}{10}}, \bibinfo{pages}{137}
  (\bibinfo{year}{2012}), \eprint{1205.6344}.

\bibitem[{\citenamefont{Peterson et~al.}(1983)\citenamefont{Peterson,
  Schlatter, Schmitt, and Zerwas}}]{Peterson:1982ak}
\bibinfo{author}{\bibfnamefont{C.}~\bibnamefont{Peterson}},
  \bibinfo{author}{\bibfnamefont{D.}~\bibnamefont{Schlatter}},
  \bibinfo{author}{\bibfnamefont{I.}~\bibnamefont{Schmitt}}, \bibnamefont{and}
  \bibinfo{author}{\bibfnamefont{P.~M.} \bibnamefont{Zerwas}},
  \bibinfo{journal}{Phys. Rev. D} \textbf{\bibinfo{volume}{27}},
  \bibinfo{pages}{105} (\bibinfo{year}{1983}).

\bibitem[{\citenamefont{Moore and Teaney}(2005)}]{Moore:2004tg}
\bibinfo{author}{\bibfnamefont{G.~D.} \bibnamefont{Moore}} \bibnamefont{and}
  \bibinfo{author}{\bibfnamefont{D.}~\bibnamefont{Teaney}},
  \bibinfo{journal}{Phys. Rev. C} \textbf{\bibinfo{volume}{71}},
  \bibinfo{pages}{064904} (\bibinfo{year}{2005}), \eprint{hep-ph/0412346}.

\bibitem[{\citenamefont{Chatrchyan et~al.}(2011)}]{Chatrchyan:2011eka}
\bibinfo{author}{\bibfnamefont{S.}~\bibnamefont{Chatrchyan}}
  \bibnamefont{et~al.} (\bibinfo{collaboration}{CMS}), \bibinfo{journal}{JHEP}
  \textbf{\bibinfo{volume}{07}}, \bibinfo{pages}{076} (\bibinfo{year}{2011}).

\bibitem[{\citenamefont{Chatrchyan et~al.}(2012)}]{Chatrchyan:2012wg}
\bibinfo{author}{\bibfnamefont{S.}~\bibnamefont{Chatrchyan}}
  \bibnamefont{et~al.} (\bibinfo{collaboration}{CMS}), \bibinfo{journal}{Eur.
  Phys. J.} \textbf{\bibinfo{volume}{C72}}, \bibinfo{pages}{2012}
  (\bibinfo{year}{2012}).

\bibitem[{\citenamefont{Acharya et~al.}(2021{\natexlab{a}})}]{ALICE:2020iug}
\bibinfo{author}{\bibfnamefont{S.}~\bibnamefont{Acharya}} \bibnamefont{et~al.}
  (\bibinfo{collaboration}{ALICE}), \bibinfo{journal}{Phys. Lett. B}
  \textbf{\bibinfo{volume}{813}}, \bibinfo{pages}{136054}
  (\bibinfo{year}{2021}{\natexlab{a}}), \eprint{2005.11131}.

\bibitem[{\citenamefont{Niemi et~al.}(2013)\citenamefont{Niemi, Denicol,
  Holopainen, and Huovinen}}]{Niemi:2012aj}
\bibinfo{author}{\bibfnamefont{H.}~\bibnamefont{Niemi}},
  \bibinfo{author}{\bibfnamefont{G.~S.} \bibnamefont{Denicol}},
  \bibinfo{author}{\bibfnamefont{H.}~\bibnamefont{Holopainen}},
  \bibnamefont{and} \bibinfo{author}{\bibfnamefont{P.}~\bibnamefont{Huovinen}},
  \bibinfo{journal}{Phys. Rev.} \textbf{\bibinfo{volume}{C87}},
  \bibinfo{pages}{054901} (\bibinfo{year}{2013}).

\bibitem[{\citenamefont{Abelev et~al.}(2013)}]{ALICE:2012vgf}
\bibinfo{author}{\bibfnamefont{B.}~\bibnamefont{Abelev}} \bibnamefont{et~al.}
  (\bibinfo{collaboration}{ALICE}), \bibinfo{journal}{Phys. Lett. B}
  \textbf{\bibinfo{volume}{719}}, \bibinfo{pages}{18} (\bibinfo{year}{2013}),
  \eprint{1205.5761}.

\bibitem[{\citenamefont{Acharya et~al.}(2019{\natexlab{b}})}]{ALICE:2019bdw}
\bibinfo{author}{\bibfnamefont{S.}~\bibnamefont{Acharya}} \bibnamefont{et~al.}
  (\bibinfo{collaboration}{ALICE}), \bibinfo{journal}{JHEP}
  \textbf{\bibinfo{volume}{09}}, \bibinfo{pages}{108}
  (\bibinfo{year}{2019}{\natexlab{b}}), \eprint{1901.05518}.

\bibitem[{\citenamefont{Heinz and Snellings}(2013)}]{Heinz:2013th}
\bibinfo{author}{\bibfnamefont{U.}~\bibnamefont{Heinz}} \bibnamefont{and}
  \bibinfo{author}{\bibfnamefont{R.}~\bibnamefont{Snellings}},
  \bibinfo{journal}{Ann. Rev. Nucl. Part. Sci.} \textbf{\bibinfo{volume}{63}},
  \bibinfo{pages}{123} (\bibinfo{year}{2013}), \eprint{1301.2826}.

\bibitem[{\citenamefont{Gardim et~al.}(2013)\citenamefont{Gardim, Grassi,
  Luzum, and Ollitrault}}]{Gardim:2012dc}
\bibinfo{author}{\bibfnamefont{F.~G.} \bibnamefont{Gardim}},
  \bibinfo{author}{\bibfnamefont{F.}~\bibnamefont{Grassi}},
  \bibinfo{author}{\bibfnamefont{M.}~\bibnamefont{Luzum}}, \bibnamefont{and}
  \bibinfo{author}{\bibfnamefont{J.-Y.} \bibnamefont{Ollitrault}},
  \bibinfo{journal}{Nucl. Phys. A} \textbf{\bibinfo{volume}{904-905}},
  \bibinfo{pages}{503c} (\bibinfo{year}{2013}), \eprint{1210.8422}.

\bibitem[{\citenamefont{Voloshin et~al.}(2010)\citenamefont{Voloshin,
  Poskanzer, and Snellings}}]{Voloshin:2008dg}
\bibinfo{author}{\bibfnamefont{S.~A.} \bibnamefont{Voloshin}},
  \bibinfo{author}{\bibfnamefont{A.~M.} \bibnamefont{Poskanzer}},
  \bibnamefont{and}
  \bibinfo{author}{\bibfnamefont{R.}~\bibnamefont{Snellings}},
  \bibinfo{journal}{Landolt-Bornstein} \textbf{\bibinfo{volume}{23}},
  \bibinfo{pages}{293} (\bibinfo{year}{2010}), \eprint{0809.2949}.

\bibitem[{\citenamefont{Mohapatra}(2016)}]{Mohapatra:2016kii}
\bibinfo{author}{\bibfnamefont{S.}~\bibnamefont{Mohapatra}},
  \bibinfo{journal}{Nucl. Phys. A} \textbf{\bibinfo{volume}{956}},
  \bibinfo{pages}{59} (\bibinfo{year}{2016}).

\bibitem[{\citenamefont{Aad et~al.}(2015)}]{ATLAS:2015qwl}
\bibinfo{author}{\bibfnamefont{G.}~\bibnamefont{Aad}} \bibnamefont{et~al.}
  (\bibinfo{collaboration}{ATLAS}), \bibinfo{journal}{Phys. Rev. C}
  \textbf{\bibinfo{volume}{92}}, \bibinfo{pages}{034903}
  (\bibinfo{year}{2015}), \eprint{1504.01289}.

\bibitem[{\citenamefont{Gardim et~al.}(2012)\citenamefont{Gardim, Grassi,
  Luzum, and Ollitrault}}]{Gardim:2011xv}
\bibinfo{author}{\bibfnamefont{F.~G.} \bibnamefont{Gardim}},
  \bibinfo{author}{\bibfnamefont{F.}~\bibnamefont{Grassi}},
  \bibinfo{author}{\bibfnamefont{M.}~\bibnamefont{Luzum}}, \bibnamefont{and}
  \bibinfo{author}{\bibfnamefont{J.-Y.} \bibnamefont{Ollitrault}},
  \bibinfo{journal}{Phys. Rev.} \textbf{\bibinfo{volume}{C85}},
  \bibinfo{pages}{024908} (\bibinfo{year}{2012}).

\bibitem[{\citenamefont{Bilandzic et~al.}(2014)\citenamefont{Bilandzic,
  Christensen, Gulbrandsen, Hansen, and Zhou}}]{Bilandzic:2013kga}
\bibinfo{author}{\bibfnamefont{A.}~\bibnamefont{Bilandzic}},
  \bibinfo{author}{\bibfnamefont{C.~H.} \bibnamefont{Christensen}},
  \bibinfo{author}{\bibfnamefont{K.}~\bibnamefont{Gulbrandsen}},
  \bibinfo{author}{\bibfnamefont{A.}~\bibnamefont{Hansen}}, \bibnamefont{and}
  \bibinfo{author}{\bibfnamefont{Y.}~\bibnamefont{Zhou}},
  \bibinfo{journal}{Phys. Rev. C} \textbf{\bibinfo{volume}{89}},
  \bibinfo{pages}{064904} (\bibinfo{year}{2014}), \eprint{1312.3572}.

\bibitem[{\citenamefont{Acharya et~al.}(2021{\natexlab{b}})}]{ALICE:2021adw}
\bibinfo{author}{\bibfnamefont{S.}~\bibnamefont{Acharya}} \bibnamefont{et~al.}
  (\bibinfo{collaboration}{ALICE}), \bibinfo{journal}{Phys. Lett. B}
  \textbf{\bibinfo{volume}{818}}, \bibinfo{pages}{136354}
  (\bibinfo{year}{2021}{\natexlab{b}}), \eprint{2102.12180}.

\end{thebibliography}

\end{document}